\documentclass[journal=jacsat,manuscript=article]{achemso}
\usepackage{chemformula} 
\usepackage[T1]{fontenc} 
\usepackage{epsfig}
\usepackage{graphicx}
\usepackage{titlesec}
\usepackage{subcaption}
\usepackage{dcolumn}
\usepackage{amsthm,amsmath}
\usepackage{comment}
\usepackage[colorlinks]{hyperref}
\usepackage{xcolor}
\usepackage[T1]{fontenc}
\usepackage{mathrsfs}
\usepackage{longtable}
\usepackage{orcidlink}
\usepackage{amsmath}
\usepackage{tabularx}
\usepackage{tabularray}
\usepackage{threeparttable}
\usepackage{caption}
\usepackage[utf8]{inputenc}
\usepackage[T1]{fontenc}
\usepackage[normalem]{ulem}

\usepackage{resizegather}
\UseRawInputEncoding

\author{Somesh Chamoli}
\affiliation[Unknown University]
{\small Department of Chemistry, Indian Institute of Technology Bombay, Powai, Mumbai 400076, India}
\author{Sudipta Chakraborty}
\affiliation[Unknown University]
{\small Department of Chemistry, Indian Institute of Technology Bombay, Powai, Mumbai 400076, India}
\author{Xubo Wang}
\affiliation[Unknown University]
{\small Department of Chemistry, the Johns Hopkins University, Baltimore, Maryland 21218, United States}
\author{Achintya Kumar Dutta}
\email{achintya@chem.iitb.ac.in}
\affiliation[Unknown University]
{\small Department of Chemistry, Indian Institute of Technology Bombay, Powai, Mumbai 400076, India}

\title[An \textsf{achemso} demo]
  {\Large Relativistic Exact-Two-Component Core-Valence-Separated Algebraic Diagrammatic Construction Theory For Near L-edge X-ray Absorption Spectra}


\begin{document}

\maketitle
\begin{abstract}
We present an efficient implementation of the second-order two-component relativistic core-valence-separated algebraic diagrammatic construction method (CVS-ADC(2)) for core-excitation calculations. The approach employs state-averaged frozen natural spinors (SA-FNS) to reduce the number of floating-point operations, together with the Cholesky decomposition (CD) technique, which lowers the storage requirements associated with two-electron integrals. These reductions make the method particularly well-suited for systems containing heavy elements. Systematic benchmarking against four-component reference calculations confirms the reliability and robustness of the two-component (X2CMP/X2CAMF)-based framework. The close agreement with canonical results further demonstrates that the SA-FNS-based CVS-ADC(2) approach achieves comparable accuracy at only a fraction of the computational cost. Moreover, benchmark studies of L$_{2,3}$-edge spectra for 3\textit{d} transition-metal compounds demonstrate that CVS-ADC(2) serves as a computationally efficient and reliable alternative to the non-Hermitian EOM-CC method for reproducing experimental spectra. Finally, calculations on a ruthenium complex illustrate the method's applicability to relativistic studies of medium-sized molecular systems. 
\end{abstract}

\section{Introduction}
X-ray absorption spectroscopy (XAS) is an advanced analytical technique that provides detailed information on the local electronic structure and atomic arrangement in molecules and materials\cite{doi:10.1021/cr9900681,Stohr1992NEXAFS,MILNE201444}. It is widely employed to probe chemical environments\cite{PhysRevA.79.053201,SOLDATOV2018232}, bonding characteristics\cite{doi:10.1021/jp411782y,BAKER2017182,D3CP03149G}, and oxidation states\cite{Cressey1993,ADAK20178,C8SC00550H}, enabling the investigation of a broad range of chemical and material properties.
Recent advances in XAS measurements have greatly broadened the range of accessible information, motivating the development of sophisticated theoretical and computational methods for reliable data interpretation\cite{doi:10.1021/acs.jctc.8b00249,doi:10.1021/acs.chemrev.8b00156,doi:10.1021/acs.accounts.0c00171,doi:10.1021/acs.jpca.0c11267,https://doi.org/10.1002/wcms.1527,doi:10.1021/acs.jctc.0c01082}.
Accurate modeling and interpretation of X-ray spectroscopic processes require the proper treatment of relativistic effects in theoretical methods, since the core-level orbitals involved in XAS are strongly influenced by relativistic phenomena. These effects are reflected in energy shifts of spectral features due to scalar relativistic contributions, along with fine-structure splitting originating from spin-orbit interactions. Such considerations are particularly important for XAS at the L- and M-edges, where spin-orbit coupling splits the \textit{p} and \textit{d} core levels into $p_{1/2}$ and $p_{3/2}$, or $d_{3/2}$ and $d_{5/2}$ sublevels. The four-component (4c) Dirac-Coulomb (DC) Hamiltonian framework is regarded as the standard theoretical approach for accurately describing scalar and spin-orbit relativistic effects. A broad range of prior studies have established the effectiveness of 4c relativistic Hamiltonians for the theoretical treatment of core-level X-ray spectroscopies, underscoring their accuracy and reliability\cite{PhysRevA.73.022501,HidekazuIkeno_2009,C5CP03712C,C6CP00561F,C6CP00262E,doi:10.1021/acs.jctc.0c01203,doi:10.1021/acs.inorgchem.1c02412}. However, 4c calculations become computationally demanding for larger molecular systems due to the high cost of evaluating, transforming, and storing two-electron integrals in the molecular spinor basis, which limits their widespread application. Consequently, two-component relativistic Hamiltonians\cite{PhysRevA.33.3742,https://doi.org/10.1002,10.1063/1.473860,NAKAJIMA1999383,BARYSZ2001181,10.1063/1.3159445,https://doi.org/10.1002/cphc.201100682} have emerged as a promising alternative, offering a favorable balance between computational efficiency and accuracy. Among the available two-component approaches, the exact two-component (X2C) theory\cite{10.1063/1.473860,10.1063/1.3159445,10.1063/1.2137315,10.1063/1.2436882,10.1093/oso/9780195140866.001.0001} stands out as a widely recognized and prominent method. Within X2C theory, several methodological variants have been developed. Among these, approaches employing one-center approximations to relativistic two-electron terms\cite{10.1063/1.5023750,doi:10.1021/acs.jpca.2c02181,10.1063/5.0095112}, as well as approaches based on model potential (MP) schemes\cite{10.1063/1.2222365,10.1063/1.2772856,10.1063/5.0095112,10.1063/5.0268348}, are particularly attractive because they avoid the evaluation of molecular relativistic two-electron integrals. This makes these approaches especially advantageous for X-ray spectroscopic applications involving large molecular systems. A wide range of theoretical X-ray spectroscopic studies based on two-component relativistic Hamiltonians has been reported in the literature.\cite{doi:10.1021/acs.jctc.7b01279,10.1063/1.5091807,PhysRevA.100.022507,D2CP00993E,10.1063/5.0300670,Wang_2025,10.1063/5.0284813}. In particular, X2C implementations employing atomic mean-field (AMF)\cite{HE1996365} integrals within density functional theory (DFT) frameworks have also been reported for X-ray absorption spectra.\cite{doi:10.1021/acs.jpca.2c08307,doi:10.1021/acs.jpclett.2c03599}. Despite the favorable computational cost of DFT, self-interaction errors can lead to significant energy shifts\cite{https://doi.org/10.1002/qua.21025}, and its accuracy depends strongly on the choice of exchange-correlation functional.\cite{doi:10.1021/acs.accounts.0c00171,doi:10.1021/acs.jctc.5b00169}. An alternative approach is provided by the equation-of-motion coupled cluster (EOM-CC) theory\cite{RevModPhys.40.153,10.1063/1.464746,10.1063/1.468592}, which offers a reliable and systematically improvable framework for treating electron correlation. EOM-CC calculations for core-excited states often suffer from convergence difficulties due to strong coupling between core-excited states and high-lying valence excited states. To address this issue, the core-valence separation (CVS) scheme\cite{PhysRevA.22.206} is commonly employed. The CVS-EOM-CC method\cite{10.1063/1.4935712} exploits the large energy separation between core and valence orbitals and explicitly excludes high-lying valence excited states from the calculation, introducing only minor errors in the calculated energies. An extensive body of literature is available on relativistic CVS-EOM-CC approaches for the description of core-excited states\cite{doi:10.1021/acs.jctc.0c01203,doi:10.1021/acs.jpclett.0c02027,https://doi.org/10.1002/wcms.1536,PhysRevA.100.022507,Wang_2025}. Within this context, an efficient CVS-EOM-CC implementation based on the X2C atomic mean-field (X2CAMF) Hamiltonian has been introduced to enable calculations of L- and M-edge core-ionized and core-excited states\cite{10.1063/5.0300670}. More recently, De-prince and co-workers have reported a CVS-EOM-CC method within the exact two-component molecular mean-field (X2CMMF) framework for the calculation of L-edge absorption spectra\cite{10.1063/5.0284813}.
Although CVS-EOM-CC methods provide a reliable and accurate description of X-ray absorption spectra, the computation of transition properties within the EOM-CC framework is challenging due to the non-Hermitian nature of the Hamiltonian, which requires the evaluation of both left and right eigenvectors. Moreover, the iterative character and high computational scaling of EOM-CC methods significantly limit their applicability to XAS calculations for large molecular systems. Another promising wavefunction-based approach that has recently attracted considerable attention is the algebraic diagrammatic construction (ADC) theory.\cite{schirmer2004intermediate,mertins1996algebraic,schirmer1991closed,schirmer1982beyond,von1984computational,dreuw2015algebraic,dempwolff2019efficient,banerjeeEfficientImplementationSinglereference2021} Owing to its Hermitian formulation, non-iterative formalism, and comparatively low computational cost, the ADC method provides an efficient and practical alternative for the calculation of X-ray absorption spectra, particularly for larger molecular systems.\cite{trofimov2000core, wenzel2014calculating, wenzel2014calculating_jcc, fransson2018simulating} ADC methods have been widely applied to the calculation of excited-state properties within the nonrelativistic framework.\cite{schirmer2004intermediate, maier2023consistent, scheurer2020complex} In contrast, their extension to the relativistic domain remains comparatively limited. Notable contributions in this area include the work of Pernpointner et al., who reported the implementation of four-component relativistic second-order and extended second-order ADC approaches for evaluating excitation energies and transition dipole moments.\cite{pernpointner2014relativistic,pernpointner2018four} Furthermore, non-Dyson relativistic ADC schemes at second and third orders have been proposed to describe valence ionization energies and electron decay phenomena.\cite{pernpointner2004one,pernpointner2010four,pernpointner2005effect, mandal2026third} A four-component third-order ADC approach for calculating ionization potentials, electron affinities, excitation energies, and excited-state properties of atoms and molecules containing heavy elements has also been reported by Dutta and co-workers.\cite{10.1063/5.0246920} Sokolov and co-workers have introduced single-reference and multireference ADC methods for ionization and electron attachment based on a one-electron variant of the X2C (X2C-1e) Hamiltonian.\cite{banerjee2019third} Despite these developments, relativistic ADC calculations targeting core-excited states remain largely unexplored, leaving significant scope for further methodological development and applications. The proven effectiveness of the CVS-ADC approach for treating core excitations within the non-relativistic framework provides strong motivation for its extension to the relativistic domain. Although the CVS-ADC scheme significantly lowers the floating-point operation count for evaluating core-level spectra in relativistic calculations, the intrinsic cost associated with storing the two-electron integrals and intermediates for medium- and large-sized systems remains a major bottleneck. To address these challenges, several cost-reduction techniques, such as density fitting\cite{10.1063/1.4807612,10.1063/1.4906344} and Cholesky decomposition\cite{HELMICHPARIS201938,10.1063/5.0161871,doi:10.1021/acs.jpca.4c04353,doi:10.1021/acs.jctc.3c01236}, aimed at reducing the storage requirements of two-electron integrals, have been adopted in relativistic calculations. Moreover, an effective strategy for reducing the computational cost of relativistic calculations involves minimizing the number of floating-point operations through the use of natural spinors\cite{10.1063/5.0085932,10.1063/5.0087243,doi:https://doi.org/10.1002/9781394217656.ch5}. MP2-based frozen natural spinors (FNS) have proven effective in reducing the computational cost of relativistic calculations across a range of wavefunction-based methods\cite{10.1063/5.0085932,10.1063/5.0125868,10.1063/5.0207091,doi:10.1021/acs.jctc.5c01791,doi:10.1021/acs.jctc.5c00199,10.1063/5.0274450,doi:10.1021/acs.jctc.5c00965,10.1063/5.0305523,thapa2026inclusionthreebodycorrectionrelativistic,10.1063/5.0305197,chakraborty2026lowcostrelativisticalgebraic}. However, they are not suitable for describing excited states, as they fail to accurately represent the corresponding electronic distributions, as demonstrated by Gomes and co-workers\cite{10.1063/5.0087243}. To address this issue, state-specific frozen natural spinors (SS-FNS) were developed\cite{10.1063/5.0289155}. Recently, a low-cost relativistic third-order ADC method for ionized, electron-attached, and excited states using CD and SS-FNS with the X2CAMF Hamiltonian has also been reported\cite{chakraborty2026lowcostrelativisticalgebraic}. While the SS-FNS approach offers a more accurate description of individual excited-state electronic distributions, the need to perform separate integral transformations for each excited state substantially reduces its efficiency, especially in the relativistic domain. To overcome this limitation, the present work employs state-averaged frozen natural spinors (SA-FNS), in which a single averaged density matrix is constructed from the targeted excited states, and the resulting natural spinors can then be used for all excited states. 
In this work, we present the theory, implementation, and benchmarking of a low-cost relativistic CVS-ADC(2) method for core-excited states using SA-FNS and CD in combination with the X2CAMF/X2CMP Hamiltonian. The ability of the implementation to model L-edge spectra of first-row transition metals is investigated, and its applicability is further demonstrated through core-excited state calculations on a medium-sized complex. 

\section{Theory}
\subsection{The X2CMP and X2CAMF schemes}
The four-component relativistic Hamiltonian in the occupation-number representation reads
\begin{equation}
\label{eq:4c}
\hat{H}^{\text{4c}}=\sum_{pq}{f_{pq}^{\text{4c}}}\,\{a_{p}^{\dagger}a_{q}\}
+\tfrac{1}{4}\sum_{pqrs}{g_{pqrs}^{\text{4c}}}\,
\{a_{p}^{\dagger}a_{q}^{\dagger}a_{s}a_{r}\},
\quad
f_{pq}^{\text{4c}}=h_{pq}^{\text{4c}}+g_{pq}^{\text{4c,MF}},
\quad
g_{pq}^{\text{4c,MF}}=\sum_{i}g_{pi,qi}^{\text{4c}},
\end{equation}
where $p,q,\ldots$ label four-component positive-energy states, the braces denote normal ordering with respect to a reference function, and $g^{\text{4c}}$ is the antisymmetrized relativistic two-electron interaction. The corresponding exact two-component (X2C) Hamiltonian, in which the fluctuation potential is approximated at the non-relativistic level, takes the analogous form
\begin{equation}
\label{eq:x2c}
\hat{H}^{\text{X2C}}=\sum_{pq}{f_{pq}^{\text{X2C}}}\,\{a_{p}^{\dagger}a_{q}\}
+\tfrac{1}{4}\sum_{pqrs}{g_{pqrs}^{\text{nr}}}\,
\{a_{p}^{\dagger}a_{q}^{\dagger}a_{s}a_{r}\},
\quad
f_{pq}^{\text{X2C}}=h_{pq}^{\text{X2C}}+g_{pq}^{\text{nr,MF}},
\quad
g_{pq}^{\text{nr,MF}}=\sum_{i}g_{pi,qi}^{\text{nr}},
\end{equation}
where $p,q,\ldots$ now label two-component spinors and $g^{\text{nr}}$ is the antisymmetrized non-relativistic Coulomb interaction. In the X2C-1e scheme, $h^{\text{X2C-1e}}$ is obtained by decoupling $h^{\text{4c}}$, and the replacement of $g^{\text{4c}}$ by $g^{\text{nr}}$ in both the mean-field and fluctuation contributions amounts to the neglect of the two-electron picture-change (2e-PC) correction.

The X2C mean-field (X2CMF) scheme recovers the 2e-PC contribution at the SCF level by decoupling the full four-component Fock matrix $f^{\text{4c}}$ built from the molecular density, at the cost of evaluating all molecular relativistic two-electron integrals. The residual picture-change correction to the fluctuation potential, i.e., the remaining difference between $g^{\text{4c}}$ and $g^{\text{nr}}$ in Eq.~\eqref{eq:x2c}, is neglected in this and all subsequent schemes considered here, but can be restored within the recently proposed X2Ccorr scheme\cite{2026WangJCTC}.

The X2C model-potential (X2CMP) scheme\cite{10.1063/5.0268348} reduces the cost of X2CMF by assembling $f^{\text{4c}}$ from a model density $\rho^{\text{MP}}$, taken as a direct sum of atomic densities, before decoupling. The 2e-PC contribution then enters as a transferable effective one-electron model potential,
\begin{equation}
\label{eq:mp}
h^{\text{MP}}=h^{\text{X2CMF}}[\rho^{\text{MP}}]-h^{\text{X2C-1e}}[\rho^{\text{MP}}],
\end{equation}
which is added to $h^{\text{X2C-1e}}$ to give $h^{\text{X2CMP}}$. The latter carries both scalar and spin-dependent 2e-PC corrections, along with any Gaunt or Breit contributions inherited from the parent four-component Hamiltonian. The construction of $h^{\text{MP}}$ requires only up to three-center relativistic two-electron integrals, and these are evaluated only once.

The X2C atomic mean-field (X2CAMF) scheme\cite{10.1063/1.5023750,doi:10.1021/acs.jpca.2c02181} is obtained from X2CMP by invoking a one-center (atomic) approximation to the remaining relativistic two-electron integrals. This approximation is well justified for the spin-dependent two-electron contributions, which are highly localized in the vicinity of the nuclei; the scalar 2e-PC correction, in contrast, originates from less localized contributions and is therefore omitted in X2CAMF. The resulting $h^{\text{X2CAMF}}$ is free of all multi-center relativistic two-electron integrals.
\subsection{Relativistic Algebraic Diagrammatic Construction Method}
Historically, the ADC framework was developed within Green's function theory, employing the conventional M{\o}ller-Plesset partitioning of the Hamiltonian. However, a unique Green's function cannot be defined for many-body systems, so propagators are introduced to treat excitation phenomena.\cite{schirmer1982beyond,dreuw2015algebraic,schirmer2018many,schirmer1991closed,oddershede1978polarization, ODDERSHEDE198433,oddershede1987propagator,schirmer1983new,ortiz2013electron} An alternative and more refined formulation of the ADC method is based on the intermediate state representation (ISR),\cite{mertins1996algebraic,dreuw2015algebraic,schirmer2018many,schirmer1991closed,mertins1996algebraicII} which naturally enables the calculation of excited-state properties and allows for extensions to open-shell formulations of the ADC matrix. Within the ISR framework, the exact excited states ($|\Psi_n\rangle$) are expressed as linear combinations of a complete basis of intermediate states ($|\tilde{\Psi}_J\rangle$). A complete manifold of correlated excited configurations is generated by acting on the exact ground-state wave function ($\Psi_0$) with excitation operators ($\hat{C}_J$), constructed from appropriate combinations of creation and annihilation operators. For excitation energies, $\hat{C}_J$ is defined as
\begin{align}
\{\hat{C}_{J}\} &= \{\hat{a}_a^{\dagger}\hat{a}_i;\hspace{1mm} \hat{a}_a^{\dagger}\hat{a}_b^{\dagger}\hat{a}_i\hat{a}_j\}; \quad \text{with } a<b,\; i<j,\; \ldots
\end{align}
and 
\begin{align}
|\Psi_{J}^{0} \rangle = \hat{C_{J}}|\Psi_{0} \rangle \quad \text{with} \quad \hat{C_{J}}|\Psi_{0} \rangle =\hat{a}_a^{\dagger}\hat{a}_i|\Psi_{0} \rangle  \quad \text{or} \quad\hat{a}_a^{\dagger}\hat{a}_b^{\dagger}\hat{a}_i\hat{a}_j|\Psi_{0} \rangle
\label{eq:ces}
\end{align}
Since these correlated excited states do not generally form an orthogonal set,\cite{mertins1996algebraic} an explicit orthogonalization procedure, such as the Gram-Schmidt orthogonalization, is required to construct a suitable basis.\cite{schirmer1991closed} As an initial step, a set of precursor states ($|\Psi_J^{\#}\rangle$) is constructed according to:
\begin{equation}
|\Psi_J^\#\rangle = |\Psi_J^0\rangle - \sum_{K\,[K]<[J]} |\tilde{\Psi}_K\rangle \langle \tilde{\Psi}_K | \Psi_J^0 \rangle ,
\end{equation}
The precursor states are subsequently transformed into an orthonormal set through symmetric orthogonalization,
\begin{equation}
|\tilde{\Psi}_I\rangle = \sum_{J,\,[J]=[I]} S_{IJ}^{-1/2}\,|\Psi_J^{\#}\rangle ,
\end{equation}
where the overlap matrix is defined as
\begin{equation}
S_{IJ} = \langle \Psi_I^{\#} | \Psi_J^{\#} \rangle .
\end{equation}
The orthonormalized intermediate states thus obtained constitute a complete basis for the construction of the ISR secular matrix \(\mathbf{M}\),
\begin{equation}
M_{IJ} = \langle \tilde{\Psi}_I | \hat{H} - E_0 | \tilde{\Psi}_J \rangle ,
\end{equation}
with \(\hat{H}\) denoting the relativistic Hamiltonian employed in this work, namely the four-component (4c), X2CAMF, or X2CMP Hamiltonian. Using matrix representation, the ISR eigenvalue problem is written as
\begin{equation}
\mathbf{M}\mathbf{X} = \mathbf{X}\boldsymbol{\Omega},
\qquad
\mathbf{X}^\dagger \mathbf{X} = \mathbf{1},
\label{eq11}
\end{equation}
where the diagonal matrix \(\boldsymbol{\Omega}\) collects the electronic excitation energies. 
The matrix \(\mathbf{M}\) is expanded in perturbative order, and truncation of this expansion at a specified order \(n\) defines the corresponding ADC(\(n\)) approximation,
\begin{equation}
\mathbf{M} = \mathbf{M}^{(0)} + \mathbf{M}^{(1)} + \mathbf{M}^{(2)} + \mathbf{M}^{(3)} + \cdots .
\label{eq12}
\end{equation}
For the current work, retaining terms up to second order yields the ADC(2) formulation.

\subsection{Properties in ISR formulation}
The ISR framework introduced above can be straightforwardly generalized to operators beyond the Hamiltonian.\cite{schirmer2004intermediate} Let us consider a generic one-particle Hermitian operator $\hat{D}$, representing a physical observable of interest. In the second-quantized formalism, this operator is expressed as
\begin{equation}
\hat{D} = \sum_{pq} d_{pq}  \left\{ c_p^\dagger c_q \right\},
\label{eq:D_op}
\end{equation}
where $d_{pq}$ are the matrix elements of $\hat{D}$ in the spinor basis $\{\chi_p\}$,
\begin{equation}
d_{pq} = \langle \chi_p | \hat{D} | \chi_q \rangle,
\label{eq:D_me}
\end{equation}
and the curly braces denote normal ordering with respect to the reference determinant.
To compute a first-order property of a given excited state $| \Psi_n^{\mathrm{ex}} \rangle$, one evaluates the corresponding expectation value of the operator $\hat{D}$:
\begin{equation}
D_{n} = \langle \Psi_n^{\mathrm{ex}} | \hat{D} | \Psi_n^{\mathrm{ex}} \rangle .
\label{eq:Dk_def}
\end{equation}
In practical implementations, it is convenient to separate the ground- and excited-state contributions. Within the ISR formalism, this is achieved by considering the shifted operator $(\hat{D} - D_0)$. The corresponding expectation value is given by
\begin{equation}
\bar{D}_n = \sum_{IJ} X_{In}^\dagger \langle \tilde{\Psi}_I | \hat{D} - D_0 | \tilde{\Psi}_J \rangle X_{Jn}
=\mathbf{X}_n^\dagger \mathbf{G} \mathbf{X}_n
\label{eq:D_shifted}
\end{equation}
where the matrix elements of $\mathbf{G}$ are defined as
\begin{equation}
G_{I J} = \langle \tilde{\Psi}_I | \hat{D} - D_0 | \tilde{\Psi}_J \rangle .
\label{eq:G_matrix}
\end{equation}
These quantities are commonly referred to as modified excited-state transition moments.
The full expectation value of the $n^{\text{th}}$ excited state is obtained by adding back the ground-state contribution,
\begin{equation}
D_0 = \langle \Psi_0 | \hat{D} | \Psi_0 \rangle ,
\label{eq:D0_def}
\end{equation}
to the shifted expectation value in Eq.~\eqref{eq:D_shifted}. This leads to
\begin{align}
D_n &= \sum_{IJ} X_{In}^\dagger \langle \tilde{\Psi}_I | \hat{D} - D_0 | \tilde{\Psi}_J \rangle X_{Jn} +
\sum_{IJ} X_{I n}^\dagger \delta_{I J} D_0 X_{J n} \\
  &= \mathbf{X}_n^\dagger \mathbf{G} \mathbf{X}_n + D_0 .
  \label{eq:D_shifted_total}
\end{align}
Here, $D_0$ corresponds to the ground-state contribution to the property.

In this manuscript, we primarily focus on XAS calculations, and transition properties play a central role in determining spectral intensities. The transition property between the ground state and the $n^{\text{th}}$ excited state is defined as
\begin{equation}
T_{0 \to n} = \langle \Psi_n^{ex} | \hat{D} - D_0 | \Psi_0 \rangle
\end{equation}
Within the ISR framework, this can be rewritten as
\begin{equation}
T_{0 \to n} = \sum_I X_{In}^\ast F_I  = \mathbf{X}_n^\dagger \mathbf{F}
\end{equation}
where
\begin{align}
F_I = \langle \tilde{\Psi}_I | \hat{D} | \Psi_0 \rangle
\end{align}
known as the modified transition moment and can be expressed in terms of second-quantized operators as
\begin{equation}
F_I = \sum_{pq} f^{I}_{pq} d_{pq}, 
\end{equation}
with
\begin{equation}
f^{I}_{pq} = \langle \tilde{\Psi}_I | c_p^\dagger c_q | \Psi_0 \rangle, 
\end{equation}
Similar to the ADC secular matrix $\mathbf{M}$, the modified transition moment matrix $\mathbf{F}$ can also be partitioned in  terms of perturbative orders as:
\begin{equation}
\mathbf{F} = \mathbf{F}^{(0)} + \mathbf{F}^{(1)} + \mathbf{F}^{(2)} + \mathbf{F}^{(3)} + \cdots .
\end{equation}

An alternate approach is to construct the ground-to-excited state transition density matrix ($\rho^{0 \to n}_{pq}$) which can be written as
\begin{equation}
\rho^{0 \to n}_{pq} = \langle \Psi_n^{ex} | c_p^\dagger c_q | \Psi_0 \rangle
= \sum_I X_{In}^\dagger \langle \tilde{\Psi}_I | c_p^\dagger c_q | \Psi_0 \rangle
= \mathbf{X}_n^\dagger \mathbf{f}_{pq}
\end{equation}
and finally, contraction of this density matrix with the operator matrix elements yields the desired transition properties:
\begin{equation}
T_{0 \to n} = \langle \Psi_n^{ex} | \hat{D} | \Psi_0 \rangle
= \sum_{pq} d_{pq} \, \rho^{0 \to n}_{pq}
\end{equation}
In the present implementation, we evaluate electronic transition dipole moments by employing the dipole moment operator $\hat{\mu}$ in place of the general operator $\hat{D}$. In this formulation, the transition density matrix $\mathbf{\boldsymbol{\rho}^{0 \to n}}$ is constructed up to second order. It is also important to recognize that transition dipole moments themselves cannot be directly measured in experiments. We therefore also compute a related observable quantity, the oscillator strength ($f$), which can be expressed as:
\begin{equation}
f_{0 \to n} = \frac{2}{3}\omega |T_{0 \to n}|^2
\end{equation}
where $\omega$ is the excitation energy obtained from solving the ADC secular matrix.

\subsection{Core-Valence Separation (CVS)}
The computation of core-excited states poses a significant challenge for standard iterative eigenvalue solvers. These states lie in the high-energy X-ray region of the electronic spectrum, meaning that all lower-lying excited states must, in principle, be computed before reaching the core-excited manifold. As a result, the computational effort increases dramatically. A straightforward solution would be to diagonalize only the subspace corresponding to core-excited configurations. However, this is formally not exact because the electronic Hamiltonian couples valence-excited and core-excited states. In practice, these couplings are typically very small due to the large energetic separation between valence and core excitations, as well as the strong spatial localization of core orbitals. This observation forms the basis of the core-valence separation (CVS) approximation\cite{PhysRevA.22.206}. Within the CVS framework, the coupling between singly core-excited states and valence- or doubly core-excited states is neglected. Consequently, the Hamiltonian matrix effectively decouples, allowing one to isolate and diagonalize only the singly core-excited block. This substantial reduction of the effective excitation space leads to significant computational savings while maintaining high accuracy for core-level spectroscopy.
Thus, the excitation operator under the CVS scheme for an excited state 
\begin{equation}
    |\Psi_{Q}^{0}\rangle = \hat{C_{Q}}|\Psi_{0} \rangle
\end{equation}
becomes
\begin{align}
\{\hat{C}_{Q}\} &= \{\hat{a}_a^{\dagger}\hat{a}_I,\hspace{1mm} \hat{a}_a^{\dagger}\hat{a}_b^{\dagger}\hat{a}_I\hat{a}_j, \ldots\}
\end{align} 
where the label \textit{I} refers to a core occupied spinor, and the label \textit{j} refers to a valence occupied spinor.

\subsection{State-Average Frozen Natural Spinors}
Natural spinors, the relativistic counterparts of L\"owdin's natural orbitals\cite{PhysRev.97.1474}, result from the diagonalization of the correlated one-body reduced density matrix (1-RDM) constructed from a spin-orbit coupled wave function. For ground-state calculations, the 1-RDM is typically constructed from the MP2 wave function. The virtual-virtual block of the MP2 1-RDM is defined as follows: 
\begin{equation}
    \label{eq:15}
D^{\text{MP2}}_{ab}= \frac{1}{2}\sum_{cij}^{} {\frac{\langle ac||ij \rangle \hspace{0.1cm}\langle ij||bc \rangle}{\varepsilon_{ij}^{ac} \hspace{0.2cm}\varepsilon_{ij}^{bc}}}
\end{equation}
However, using MP2-based natural spinors alone to describe excited states does not lead to an accurate representation of the electronic distribution. Consequently, a method that provides a more reliable first-order description of the excited-state wave function is required. To provide a reasonably accurate description of excited states at moderate computational cost, CIS(D) calculations are first performed, after which a state-averaged (SA) approach is applied to construct a single one-body reduced density matrix (SA-1-RDM) by averaging over the selected excited states. The virtual-virtual block of the SA-1-RDM is defined as: 
\begin{equation}
    \label{eq:16}
D^{\text{SA}}_{ab}= D^{\text{MP2}}_{ab} + \frac{1}{N_{roots}}\sum_{\kappa=1}^{N_{roots}} D^{\text{CIS(D)}}_{ab}(\kappa)
\end{equation}
In Eq. (\ref{eq:16}), $D^{\text{CIS(D)}}_{ab}(\kappa)$ is the virtual-virtual block of  CIS(D) 1-RDM for the excited state $\kappa$. The explicit expressions for the CIS(D) 1-RDM are provided in the Supporting Information. The following steps are then carried out to obtain the state-averaged natural spinors. 
\begin{enumerate}
    \item Diagonalization of $D^{SA}_{ab}$ to obtain the virtual natural spinors (\textit{V}) as eigenvectors and their associated occupation numbers (\textit{n}) as eigenvalues.
    \begin{equation}
    \label{eq:17}
    D^{\text{SA}}_{ab}V=Vn
    \end{equation}
    \item The virtual-virtual block of the Fock matrix ($F_{vv}$) is transformed into the truncated natural spinor basis ($F^{\text{NS}}_{vv}$) by applying a predefined occupation-number threshold.
    \begin{equation}
    \label{eq:18}
    F^{\text{NS}}_{vv}=\Tilde{V}^{\dagger}F_{vv}\Tilde{V}
    \end{equation}
    where $\Tilde{V}$ are virtual natural spinors in a truncated basis.
    \item The $F_{vv}^{\text{NS}}$ is then diagonalized to produce the semi-canonical virtual natural spinors ($\tilde{Z}$) and the associated orbital energies ($\epsilon$).
    \begin{equation}
    \label{eq:19}
    F_{vv}^{\text{NS}}\tilde{Z}=\tilde{Z}\epsilon
    \end{equation}
    \item The transformation matrix ($U$) is constructed to convert the canonical virtual spinor space into the semi-canonical natural virtual spinor space.
    \begin{equation}
    \label{eq:20}
    U=\tilde{V}\tilde{Z}
    \end{equation}
\end{enumerate}
The resulting set of spinors is also referred to as the State Average Frozen Natural Spinors (SA-FNS), since the occupied spinors are retained in their canonical form, while only the virtual spinors are transformed into semi-canonical natural virtual spinors. Once the SA-FNS are generated, subsequent CVS-ADC(2) calculations are carried out in the SA-FNS basis. To obtain the perturbative correction associated with the truncated virtual space, CIS(D) calculations are recomputed in the SA-FNS basis. For a given state $\kappa$, this correction is defined as follows: 
\begin{equation}
    \label{eq:21}
\Delta E_{\text{CIS(D)}}(\kappa)= E^{\text{canonical}}_{\text{CIS(D)}}(\kappa)-E^{\text{SA-FNS}}_{\text{CIS(D)}}(\kappa)
\end{equation}
This correction is then used as an approximation to $\Delta E_{\text{CVS-ADC(2)}}(\kappa)$ and is added to the uncorrected CVS-ADC(2) energy to obtain the corrected value. 
\begin{equation}
    \label{eq:22}
E^{\text{corrected}}_{\text{CVS-ADC(2)}}(\kappa)= E^{\text{uncorrected}}_{\text{CVS-ADC(2)}}(\kappa) + \Delta E_{\text{CIS(D)}}(\kappa)
\end{equation}
An analogous correction for the oscillator strengths ($\Theta$) can be defined as:
\begin{equation}
    \label{eq:23}
\Delta \Theta_{\text{CIS}}(\kappa)= \Theta^{\text{canonical}}_{\text{CIS}}(\kappa)-\Theta^{\text{SA-FNS}}_{\text{CIS}}(\kappa)
\end{equation}
However, its contribution is very small and is neglected in practice. 

\section{Implementation and Computational Details}
The CVS-ADC(2) method, based on the SA-FNS framework, CD, and the X2CMP/X2CAMF Hamiltonian, has been implemented in the development version of BAGH\cite{duttaBAGHQuantumChemistry2025}. The X2CAMF-HF and X2CMP-HF calculations are performed using the \texttt{socutils} package\cite{socutils}, interfaced with BAGH.
The implementation begins with the generation of Cholesky vectors in the atomic orbital (AO) basis, followed by their transformation to the molecular orbital (MO) basis and the construction of antisymmetrized two-particle (0-2 external) integrals. Integrals involving more than two particle indices are neither constructed nor stored; instead, they are generated on the fly from the Cholesky vectors in both the canonical and SA-FNS bases.
Subsequently, the ground-state 1-RDM is constructed from MP2 amplitudes, followed by CIS and CIS(D) calculations to obtain the state-averaged 1-RDM (SA-1-RDM). The SA-1-RDM is then diagonalized to generate the state-averaged frozen natural spinors (SA-FNS).
In the next step, Cholesky vectors are transformed into the truncated SA-FNS basis, and the corresponding antisymmetrized 0-2 particle integrals are constructed. CIS and CIS(D) calculations are then repeated in the truncated SA-FNS basis to account for perturbative corrections arising from the reduced space.
Finally, CVS-ADC(2) calculations are performed in the truncated SA-FNS basis. The canonical CIS eigenvectors, transformed into the SA-FNS basis, are used as initial guesses for the subsequent CVS-ADC(2) calculations. A root-specific Davidson eigensolver\cite{HIRAO1982246} is employed to ensure that each root in the SA-FNS basis corresponds to its canonical counterpart. 

\noindent To evaluate the performance of the CVS-ADC(2) method, L$_{2,3}$ -edge X-ray absorption spectra are computed for 3\textit{d} transition metal oxyanions, MO$_{4}^{n-}$ (M = Ti, V, Cr, Mn), as well as for two additional 3\textit{d} transition metal complexes, TiCl$_{4}$ and VOCl$_{3}$. In addition, the practical applicability of the method is illustrated through the calculation of a core-excitation energy for a medium-sized metal complex. The molecular geometries of SiCl$_{4}$
and TiCl$_{4}$ were taken from Ref \cite{doi:10.1021/acs.jctc.7b01279}, while that of VOCl$_{3}$ was obtained from Ref \cite{10.1063/1.5091807}. The geometries of the metal oxyanions were taken from Ref \cite{D5CP01656H}. All calculations were performed within the frozen-core approximation, in which core spinors lying below the metal 2\textit{p} spinors are excluded from the correlation treatment. All computed spectra were broadened using a Lorentzian function, while the experimental spectra were digitized using WebPlotDigitizer\cite{WebPlotDigitizer}.

\section{Results and Discussion}
\subsection{Comparison with the Four-component Method}
\begin{figure}[h!]
    \begin{center}
    \includegraphics[width=0.7\textwidth]{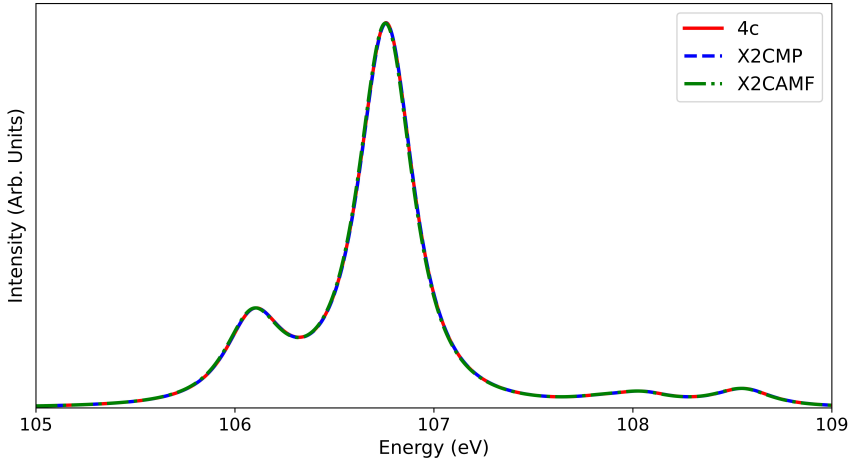}
    \caption{CVS-ADC(2) XAS spectra near Si L$_{2,3}$ -edge of SiCl$_{4}$ molecule.}
    \label{fig:my_label1}
    \end{center}
\end{figure}

Scalar two-electron picture-change (2e-PC) corrections to molecular properties have previously been shown to be small\cite{10.1063/1.2133731,10.1063/1.1904589}. For core-level excitations, particularly at the L-edge, these contributions could in principle become more significant. Within the X2CAMF framework, however, the scalar 2e-PC terms are neglected, which raises the question of whether this approximation remains reliable for L-edge core-excitation spectra. 
To address this issue, we perform a direct comparison of X-ray absorption spectra (XAS) at the L$_{2,3}$ -edge computed using the X2CAMF and X2CMP Hamiltonians with reference results obtained from the four-component CVS-ADC(2) method for the SiCl$_{4}$ molecule and the argon (Ar) atom. 
For Si, the dyall.v2z basis set was used, and for Cl, the uncontracted aug-cc-pVDZ basis set was employed, while the uncontracted 6-311(2+,+)G(p,d) basis set, supplemented by Rydberg-type functions as outlined in Ref\cite{doi:10.1021/acs.jpclett.0c02027} was used for the Ar atom. Figure \ref{fig:my_label1} compares the L$_{2,3}$ edge spectrum of SiCl$_{4}$ molecule computed with the CVS-ADC(2) method using the 4c, X2CAMF, and X2CMP Hamiltonians. 
From Figure \ref{fig:my_label1}, it is evident that the X2CAMF and X2CMP approaches yield spectra in very good agreement with the 4c results. The relative positions of the L$_{3}$ and L$_{2}$ peaks, as well as their intensities, are well reproduced by both two-component methods. The slightly closer agreement of the X2CMP spectra with the 4c reference is expected, as X2CMP explicitly includes two-electron picture-change contributions. Nevertheless, the ability of X2CAMF to accurately reproduce both peak positions and relative intensities indicates that scalar two-electron picture-change effects are relatively small even for L-edge excitations, with spin-orbit coupling being the dominant factor governing the L$_{2,3}$-edge spectral features. 

\begin{figure}[H]
    \begin{center}
    \includegraphics[width=0.7\textwidth]{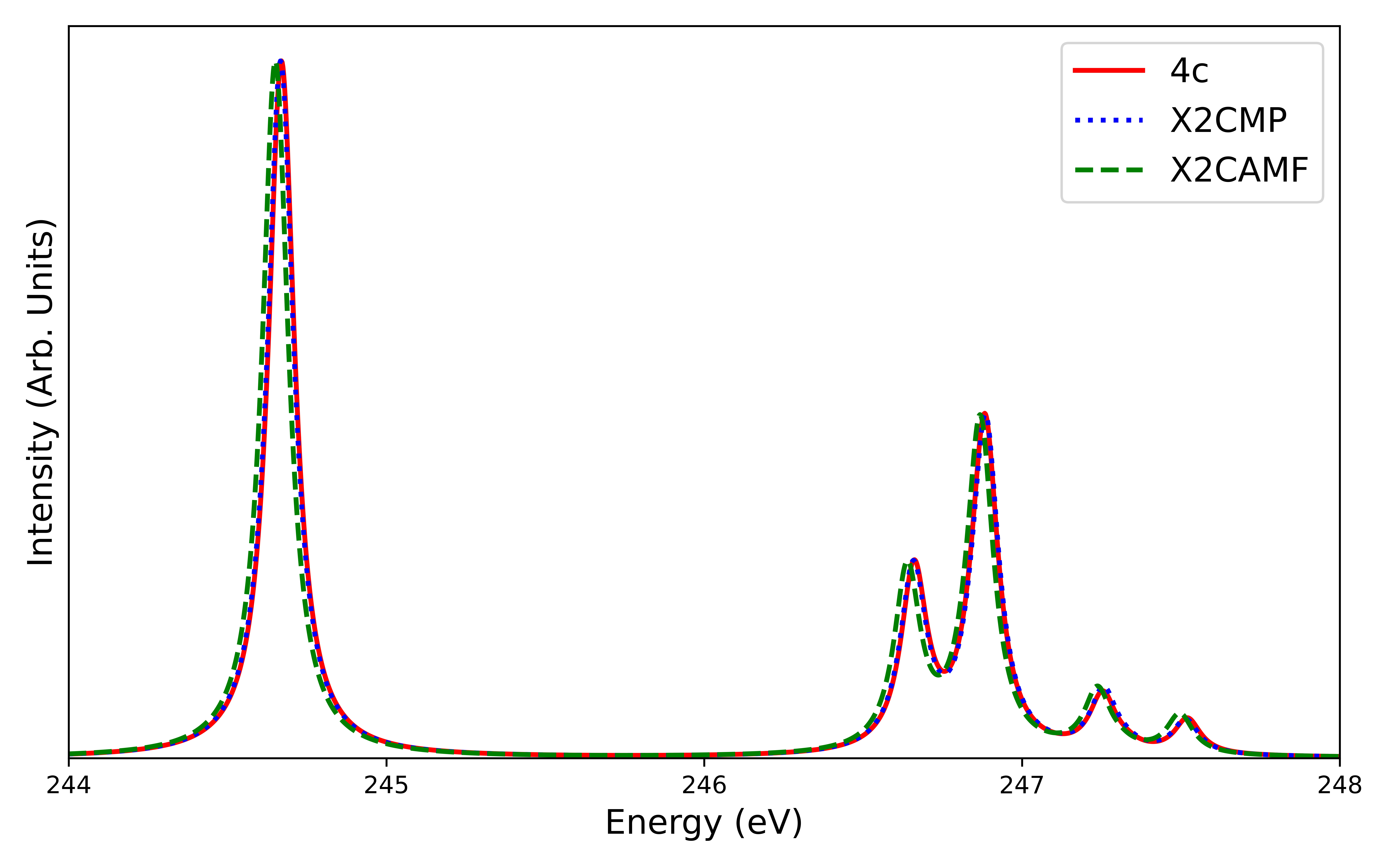}
    \caption{CVS-ADC(2) XAS spectra near Ar L$_{2,3}$ -edge for Ar atom.}
    \label{fig:my_label2}
    \end{center}
\end{figure}

\noindent An analogous analysis was performed for the Ar atom to further assess the performance of the two-component Hamiltonians for L$_{2,3}$-edge spectra (see Figure \ref{fig:my_label2}). The figure displays the first four spectral bands, corresponding to the $2p_{3/2} \xrightarrow{} 4s$, $2p_{1/2} \xrightarrow{} 4s$, $2p_{3/2} \xrightarrow{} 5s/3d$, and $2p_{3/2} \xrightarrow{} 6s/4d$ transitions, respectively. As shown in Figure \ref{fig:my_label2}, both the X2CAMF and X2CMP methods yield spectra in very good agreement with the 4c results.
\noindent The relative peak positions and their intensities are well reproduced by both two-component approaches, closely matching the 4c reference. The maximum deviation in peak positions with respect to the 4c results is 0.02 eV for X2CAMF and is further reduced to 0.01 eV for X2CMP. Furthermore, the spectral features obtained from CVS-ADC(2) are largely consistent with those from EOM-CC calculations\cite{doi:10.1021/acs.jpclett.0c02027,10.1063/5.0229955,10.1063/5.0284813}, with a slight improvement in capturing the inversion of peak intensities between the second and third peaks. The underestimation of the intensities of the third and fourth peaks may be attributed to the absence of steep d- and f-type functions in the employed basis set, as suggested by Cheng and co-workers\cite{10.1063/5.0300670}. However, these effects were not considered in the present study. 

\noindent Overall, these comparisons demonstrate that the two-component methods based on the X2CAMF and X2CMP Hamiltonians provide a robust, efficient, and accurate alternative to more computationally demanding 4c calculations for L-edge spectra. Due to its slightly better agreement with the 4c results, the X2CMP Hamiltonian is adopted for all subsequent calculations.

\subsection{Effect of higher-order relativistic effects}
\begin{figure}[H]
    \begin{center}
    \includegraphics[width=0.7\textwidth]{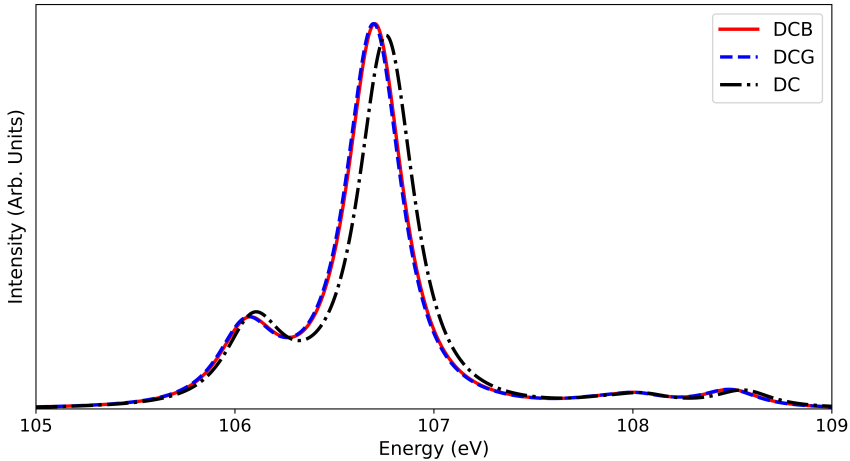}
    \caption{Effect of higher-order relativistic effects on calculated CVS-ADC(2) XAS spectra near Si L$_{2,3}$ -edge of SiCl$_{4}$ molecule.}
    \label{fig:my_label3}
    \end{center}
\end{figure}

To assess the impact of higher-order relativistic effects on L-edge XAS spectra within the X2CMP framework, calculations were performed with three different Hamiltonians: the Dirac-Coulomb (DC) Hamiltonian, the Dirac-Coulomb-Gaunt (DCG) Hamiltonian, and the Dirac-Coulomb-Breit (DCB) Hamiltonian for the Si L$_{2,3}$ -edge spectrum of the SiCl$_{4}$. As evident from Figure \ref{fig:my_label3}, the inclusion of the Gaunt term leads to a small but noticeable red shift of the transition energies relative to the Dirac-Coulomb reference. When the full Breit interaction is included, the resulting spectrum is virtually indistinguishable from the DCG result. This indicates that the dominant correction beyond the Coulomb term originates from the Gaunt contribution, while the remaining retardation part of the Breit operator has a negligible influence on the L$_{2,3}$ -edge excitations of SiCl$_{4}$. This confirms that such higher-order relativistic effects are minor but systematic, manifesting primarily as a slight energetic red shift without substantially altering spectral shapes or relative intensities. For consistency and to incorporate these systematic higher-order contributions, the Breit interaction was retained in all subsequent calculations.

\subsection{Convergence with Respect to the Threshold}
\begin{figure}[H]
    \begin{center}
    \includegraphics[width=0.7\textwidth]{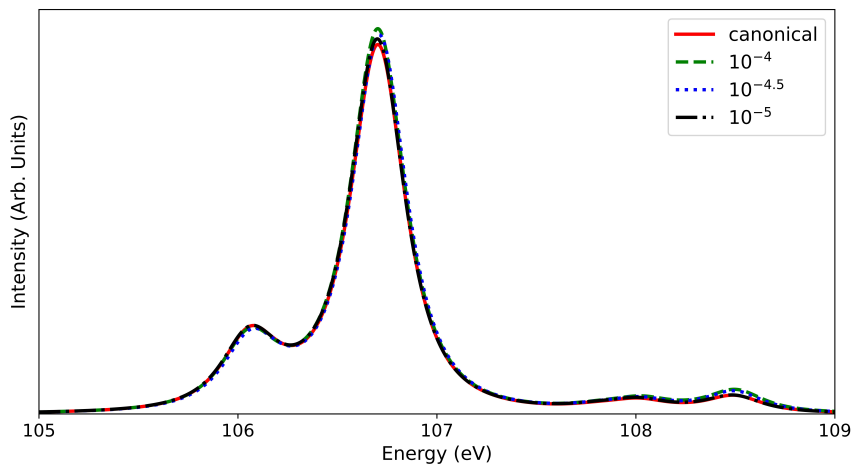}
    \caption{CVS-ADC(2) XAS spectra near Si L$_{2,3}$ -edge of SiCl$_{4}$ molecule in different truncation thresholds.}
    \label{fig:my_label4}
    \end{center}
\end{figure}

\begin{figure}[H]
    \begin{center}
    \includegraphics[width=0.7\textwidth]{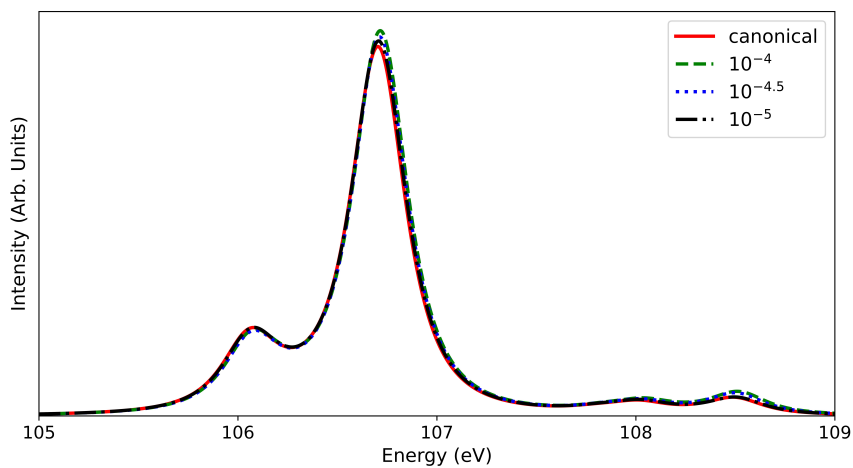}
    \caption{CVS-ADC(2) XAS spectra near Si L$_{2,3}$ -edge of SiCl$_{4}$ molecule in different truncation thresholds with correction included.}
    \label{fig:my_label5}
    \end{center}
\end{figure}

To assess the impact of the truncation threshold on computational efficiency and accuracy, three SA-FNS truncation thresholds (10$^{-4}$, 10$^{-4.5}$, and 10$^{-5}$) were considered for the SiCl$_{4}$ molecule. The dayll.v2z basis set was used for Si, and the uncontracted aug-cc-pVDZ basis set was employed for Cl. Since a Cholesky decomposition threshold of 10$^{-5}$ closely reproduces the standard integral results, it was used throughout. Canonical CVS-ADC(2) calculations with the full virtual space and a Cholesky threshold of 10$^{-5}$ were taken as the reference. Figure \ref{fig:my_label4} compares the Si L$_{2,3}$-edge spectra of the SiCl$_{4}$ computed with the CVS-ADC(2) method at three different SA-FNS truncation thresholds, alongside the canonical reference. As shown in Figure \ref{fig:my_label4}, the relative positions of the L$_{3}$  and L$_{2}$  peaks, as well as their energy separation, are accurately reproduced across all three truncation thresholds relative to the canonical spectrum. This indicates that the essential spectral features are largely insensitive to the level of truncation employed in the SA-FNS procedure. Small but noticeable differences in peak intensities are observed, as evident in the zoomed region (see Fig. S1). As the truncation threshold is tightened, the peak heights progressively converge toward those of the canonical reference, reflecting a systematic improvement in quantitative agreement. The inclusion of perturbative corrections described in Eqs.~(\ref{eq:21})--(\ref{eq:22}) leads to only minor further changes in the spectral profiles, as illustrated in Figure \ref{fig:my_label5}, where the corrected spectra at the three truncation thresholds are in even closer agreement with the canonical reference. However, this slight improvement in intensity comes at the cost of a reduced degree of truncation of the virtual space. While the 10$^{-4}$ threshold yields a substantial reduction from 400 to 166 virtual spinors (58\% truncation), tightening the threshold to 10$^{-4.5}$ and 10$^{-5}$ increases the retained virtual space to 192 (52\% truncation) and 216 (46\% truncation), respectively. Thus, although the 10$^{-5}$ threshold provides the closest agreement with the canonical intensities, it also significantly diminishes the computational savings. 

\noindent Taken together, these results suggest that a truncation threshold of 10$^{-4.5}$ offers an optimal balance between computational efficiency and spectral accuracy, retaining key spectral features and achieving near-converged intensities while substantially reducing the size of the virtual space. This makes it a practical and well-justified choice for rapid estimation of core-level spectra in heavy-element systems. 

\subsection{Benchmark Calculations}
\subsubsection{3\textit{d} Transition Metal Oxyanions}
\begin{figure}[t]
    \centering

    \begin{subfigure}{0.49\textwidth}
        \centering
        \includegraphics[width=\textwidth]{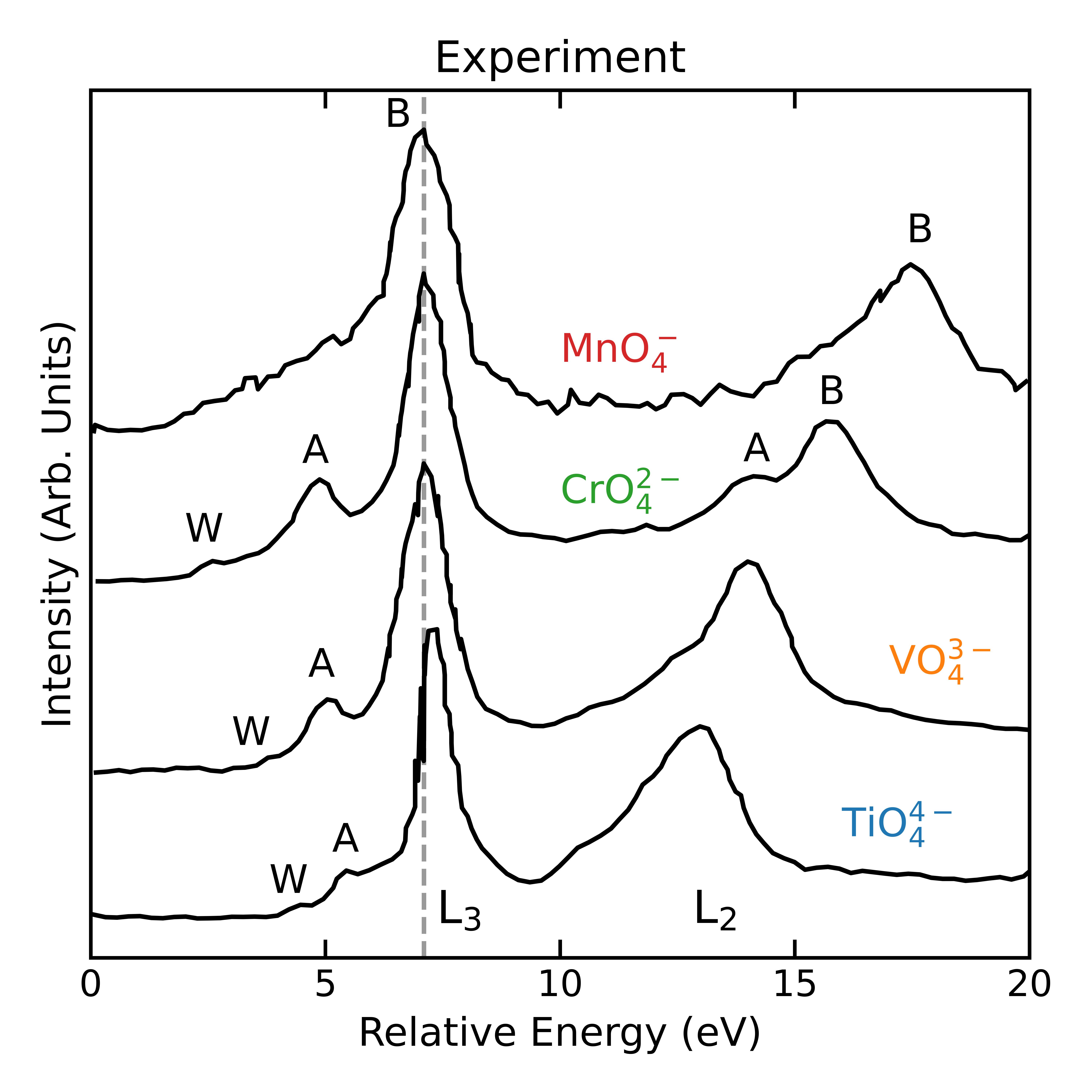}
        \caption{}
    \end{subfigure}
    \hfill
    \begin{subfigure}{0.49\textwidth}
        \centering
        \includegraphics[width=\textwidth]{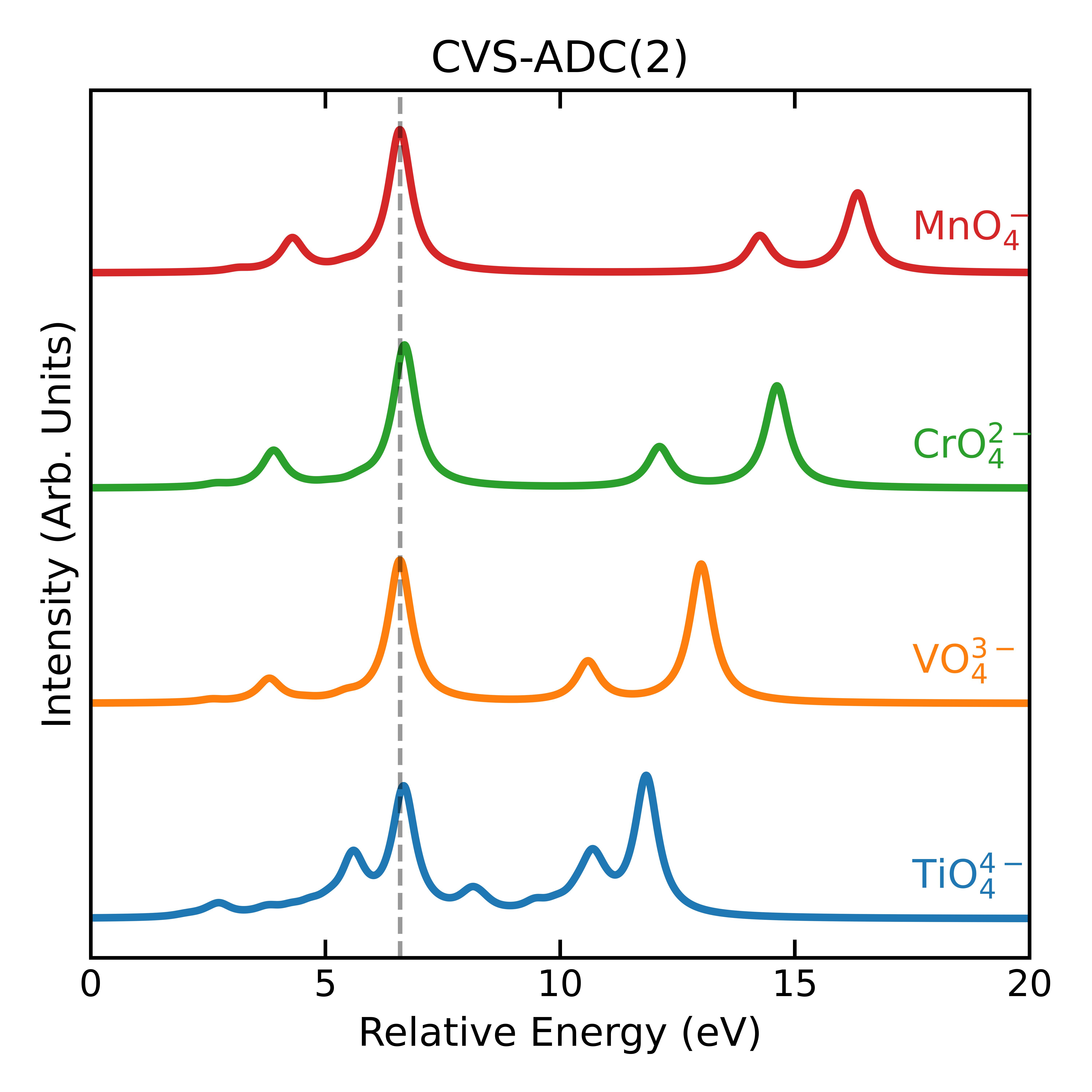} 
        \caption{}
    \end{subfigure}

    \caption{XAS spectra near M L$_{2,3}$ -edge of MO$_{4}^{n-}$ transition metal oxyanions obtained using (a) experimental method and (b) CVS-ADC(2) method.}
    \label{fig:my_label6}
\end{figure}

To assess the performance of the CVS-ADC(2) method, we first calculate the L$_{2,3}$ -edge spectra of 3\textit{d} transition metal oxyanions, MO$_{4}^{n-}$ (M = Ti, V, Cr, Mn). An uncontracted x2c-TZVPall-2c basis set was employed for all oxyanions.  
Figure \ref{fig:my_label6} compares the experimental L$_{2,3}$ -edge spectra of these metal oxyanions with those obtained from the CVS-ADC(2) method. The experimental data were taken from the work of Brydson et al.\cite{R_Brydson_1993} and digitized using WebPlotDigitizer. The CVS-ADC(2) spectra were broadened using a Lorentzian function with a full width at half maximum (FWHM) of 0.3 eV. To facilitate comparison with the experimental spectra, a uniform energy shift was applied to the computed spectra to align the L$_{3}$ -edge (marked by dashed vertical lines), thereby bringing all spectra onto a common energy scale. Figure \ref{fig:my_label6} highlights the comparison of the spin-orbit splitting between the L$_{3}$ and L$_{2}$ edges. The computed splittings follow the trend MnO$_{4}^{-}$ > CrO$_{4}^{2-}$ > VO$_{4}^{3-}$ > TiO$_{4}^{4-}$, in good agreement with the experimental observations. Furthermore, the L$_{3}$/L$_{2}$ intensity ratio follows the order MnO$_{4}^{-}$ > CrO$_{4}^{2-}$ > VO$_{4}^{3-}$ > TiO$_{4}^{4-}$, consistent with the experimental study, and this trend is also well reproduced by the computed spectra. Notably, this intensity ratio trend contrasts with that obtained using the DFT/CIS method\cite{D5CP01656H}, which predicts an inversion of peak intensities between the L$_{2}$ and L$_{3}$ edges. A direct comparison between the CVS-ADC(2) spectra and those computed using the DFT/CIS approach for the metal oxyanions is provided in the SI. In addition, the absolute L$_{2}$-L$_{3}$ splittings and L$_{3}$/L$_{2}$ intensity ratios for both the experimental and CVS-ADC(2) spectra are also reported in the SI. 

\noindent Another feature of interest is the splitting of both the L$_{3}$ and L$_{2}$ edges into two components (labeled A and B in the experimental spectra). This splitting arises from ligand-field effects, which split the unoccupied 3\textit{d} orbitals of the metal centers into \textit{e} and $t_{2}$ sets. The A feature exhibits lower relative intensity than B. The overall trend in the A-B splitting magnitude across the oxyanions is expected to follow MnO$_{4}^{-}$ > CrO$_{4}^{2-}$ > VO$_{4}^{3-}$ > TiO$_{4}^{4-}$, driven by the increase in average M-O bond lengths and the corresponding decrease in the formal charge on the transition metal center. The computed spectra reproduce this trend qualitatively, showing a slight increase in the A-B splitting for CrO$_{4}^{2-}$, consistent with the experimental observations. However, for VO$_{4}^{3-}$, the splitting remains nearly constant rather than decreasing, indicating a slight overestimation. The trend is then recovered for TiO$_{4}^{4-}$, where the splitting decreases as expected. The computed spectra also capture weak pre-edge features (W) preceding the L$_{3}$ edge in the oxyanions. 

\subsubsection{TiCl\texorpdfstring{\textsubscript{4}}{4} and VOCl\texorpdfstring{\textsubscript{3}}{3}}
\begin{figure}[t]
    \begin{center}
    \includegraphics[width=0.7\textwidth]{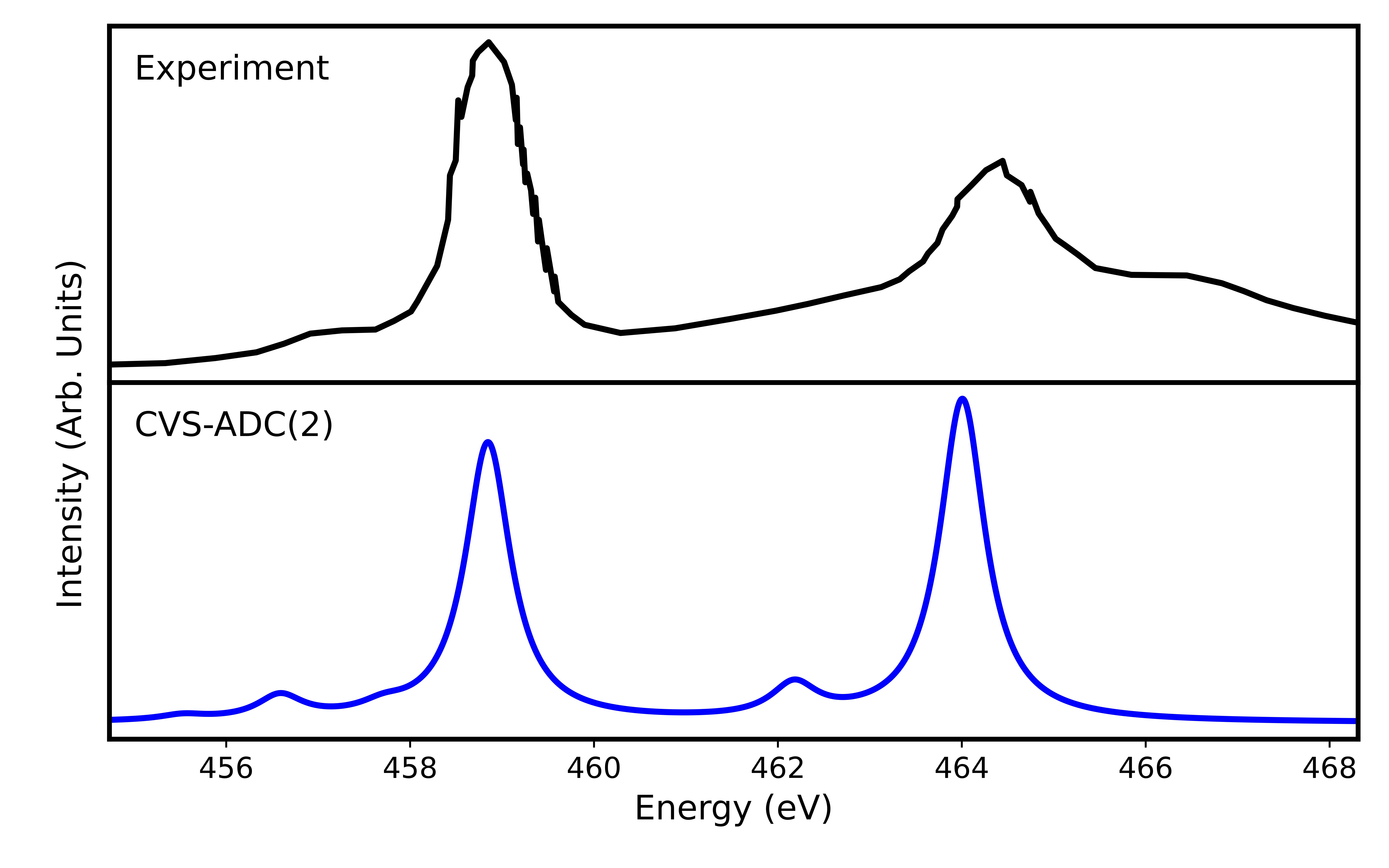}
    \caption{CVS-ADC(2) XAS spectra near Ti L$_{2,3}$ -edge of TiCl$_{4}$ and its comparison with experimental spectra.}
    \label{fig:my_label7}
    \end{center}
\end{figure}

As a further test of the CVS-ADC(2) method, we extend our benchmarking to transition metal complexes, specifically TiCl$_{4}$ and VOCl$_{3}$, two well-studied benchmark systems. An uncontracted x2c-TZVPall-2c basis set was used for both complexes. Figure \ref{fig:my_label7} presents a comparison between the experimental Ti L$_{2,3}$-edge spectrum of TiCl$_{4}$ and the spectrum computed using the CVS-ADC(2) method. The experimental spectrum was digitized from the work of Hitchcock and co-workers\cite{doi:10.1139/v93-204}, and the computed spectrum was broadened using a Lorentzian function with a FWHM of 0.3 eV. 

\noindent Like the metal oxyanions, TiCl$_{4}$ exhibits two distinct spectral features associated with the L$_{3}$ and L$_{2}$ edges, originating from Ti 2\textit{p} $\xrightarrow{}$ 3\textit{d} transitions. Additionally, subtle shoulder features appear at both edges, arising from ligand-field splitting of the Ti 3\textit{d} orbitals into \textit{e} and $t_{2}$ levels. A clear difference between the computed and experimental spectra is observed in the relative intensities of the L$_{2}$ and L$_{3}$ edges. Experimentally, the L$_{3}$ edge is more intense than the L$_{2}$ edge, whereas the CVS-ADC(2) calculations predict the opposite trend, with the L$_{2}$ edge appearing more intense. This discrepancy is not unique to the present work and has also been reported in real-time TD-DFT\cite{doi:10.1021/acs.jctc.7b01279}, DFT/CIS\cite{D5CP01656H}, and CVS-EOM-CC\cite{10.1063/5.0284813} studies of the same spectrum. Quantitatively, the experimental L$_{2}$/L$_{3}$ intensity ratio is approximately 0.7, while the CVS-ADC(2) results yield a ratio of 1.1, consistent with earlier CVS-EOM-CC calculations that report a similar value. The CVS-ADC(2) results are also consistent with EOM-CC and TD-DFT calculations in terms of the L$_{2}$-L$_{3}$ splitting, as all methods underestimate the experimental value by 0.5 eV. However, notable differences are observed in the magnitude of the energy shifts required to align the calculated spectra with the experimental spectrum: TD-DFT requires 9.5 eV, CVS-EOM-CC -2.7 eV, and CVS-ADC(2) -4.1 eV. Overall, for TiCl$_{4}$, CVS-ADC(2) exhibits trends similar to EOM-CC and TD-DFT and shows an intermediate energy shift between the two methods. 
\newline
Compared to TiCl$_{4}$, VOCl$_{3}$ shows a densely packed series of transitions in its L-edge spectrum, extending across both the L$_{3}$ and L$_{2}$ regions. This arises from the lower molecular symmetry (C$_{3v}$) of VOCl$_{3}$, which causes the V 3$d$ orbitals to split into three distinct virtual valence levels. Figure 8 compares the experimental V L$_{2,3}$-edge spectrum of VOCl$_{3}$ with the spectrum calculated using the CVS-ADC(2) method. The experimental data were digitized from the work of Fronzoni et al.\cite{doi:10.1021/jp808720z}, while the computed spectrum was broadened using a Lorentzian function with a FWHM of 0.25 eV. The computed spectrum shows reasonable qualitative agreement with the experimental data, particularly in the L$_{3}$-edge region. The overall pattern and relative intensities are well reproduced; however, the spacing between the L$_{3}$-edge peaks is overestimated, consistent with the trend observed in CVS-EOM-CC\cite{10.1063/5.0284813} calculations. The method also captures a weak pre-edge feature appearing just before the L$_{3}$ edge. 
\begin{figure}[t]
    \begin{center}
    \includegraphics[width=0.7\textwidth]{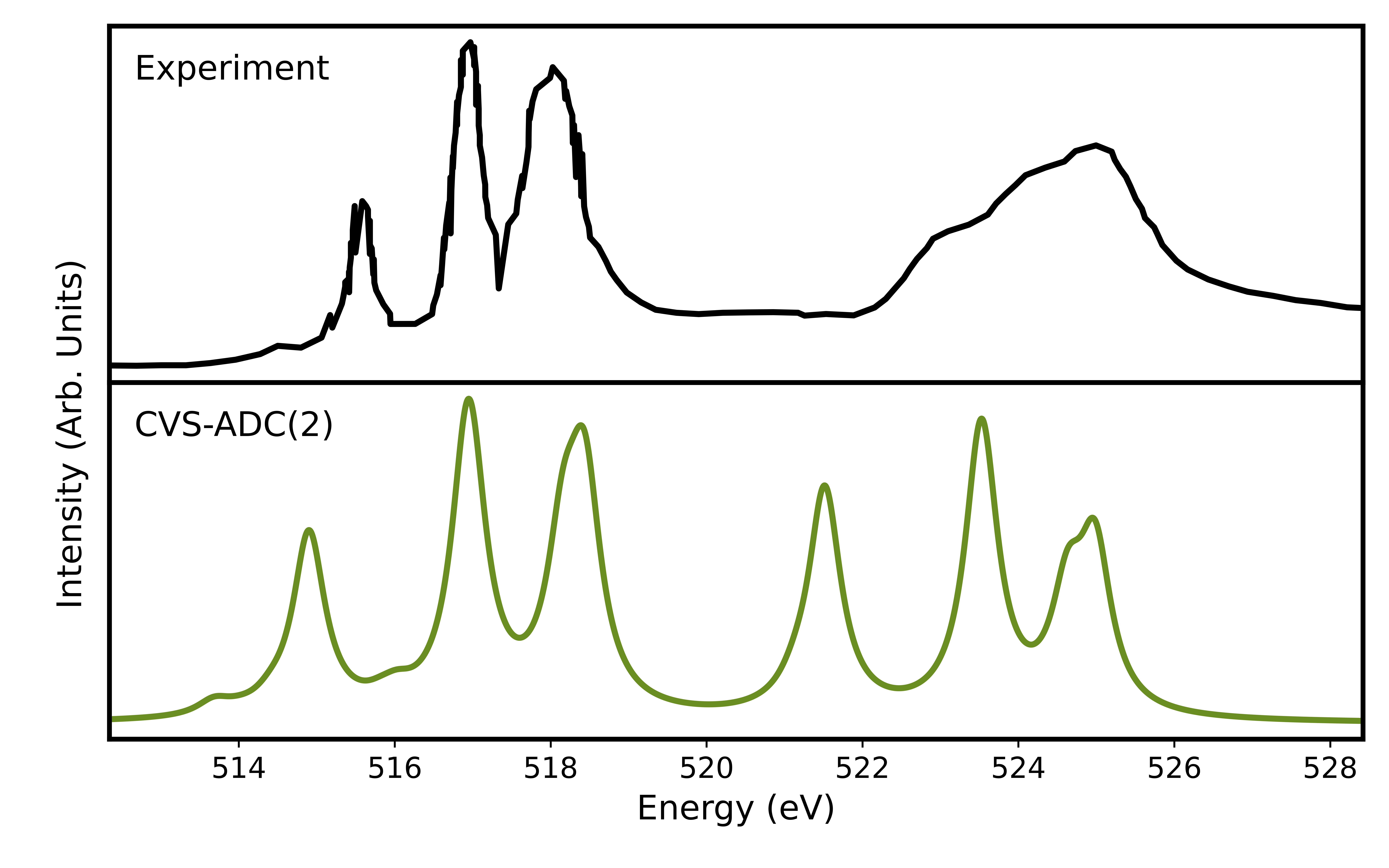}
    \caption{CVS-ADC(2) XAS spectra near V L$_{2,3}$ -edge of VOCl$_{3}$ and its comparison with experimental spectra.}
    \label{fig:my_label8}
    \end{center}
\end{figure}
This agreement, however, deteriorates in the higher-energy L$_{2}$ region, where, in line with CVS-EOM-CC results, the method shows poor agreement with experiment. In particular, the separation between the most intense features of the L$_{3}$ and L$_{2}$ edges is underestimated: against an experimental value of 7.1 eV, both CVS-EOM-CC and CVS-ADC(2) predict a smaller separation of about 6.5 eV. Moreover, the relative intensities of the peaks in the L$_{2}$ region are not reproduced in accordance with the experimental trend. A similar trend in the L$_{2}$ edge is also observed in the TDDFT study\cite{10.1063/1.5091807}. In terms of the relative intensities of the L$_{2}$ and L$_{3}$ edges, the experimental ratio is 0.7. CVS-EOM-CC overestimates this value at 1.0, whereas CVS-ADC(2) provides a somewhat improved estimate of 0.9. However, the energy shift required for CVS-EOM-CC (-3.0 eV) is less pronounced than that needed for CVS-ADC(2) (-4.7 eV). 

\noindent Based on the benchmark calculations presented above, the CVS-ADC(2) method yields L-edge spectra of comparable qualitative accuracy to the CVS-EOM-CC method, but at a substantially reduced computational cost. It therefore serves as an efficient and practical alternative for evaluating L-edge spectra, particularly in systems containing heavy elements.

\subsection{Application to Medium-Sized Complex}
\begin{figure}[H]
    \begin{center}
    \includegraphics[width=0.6\textwidth]{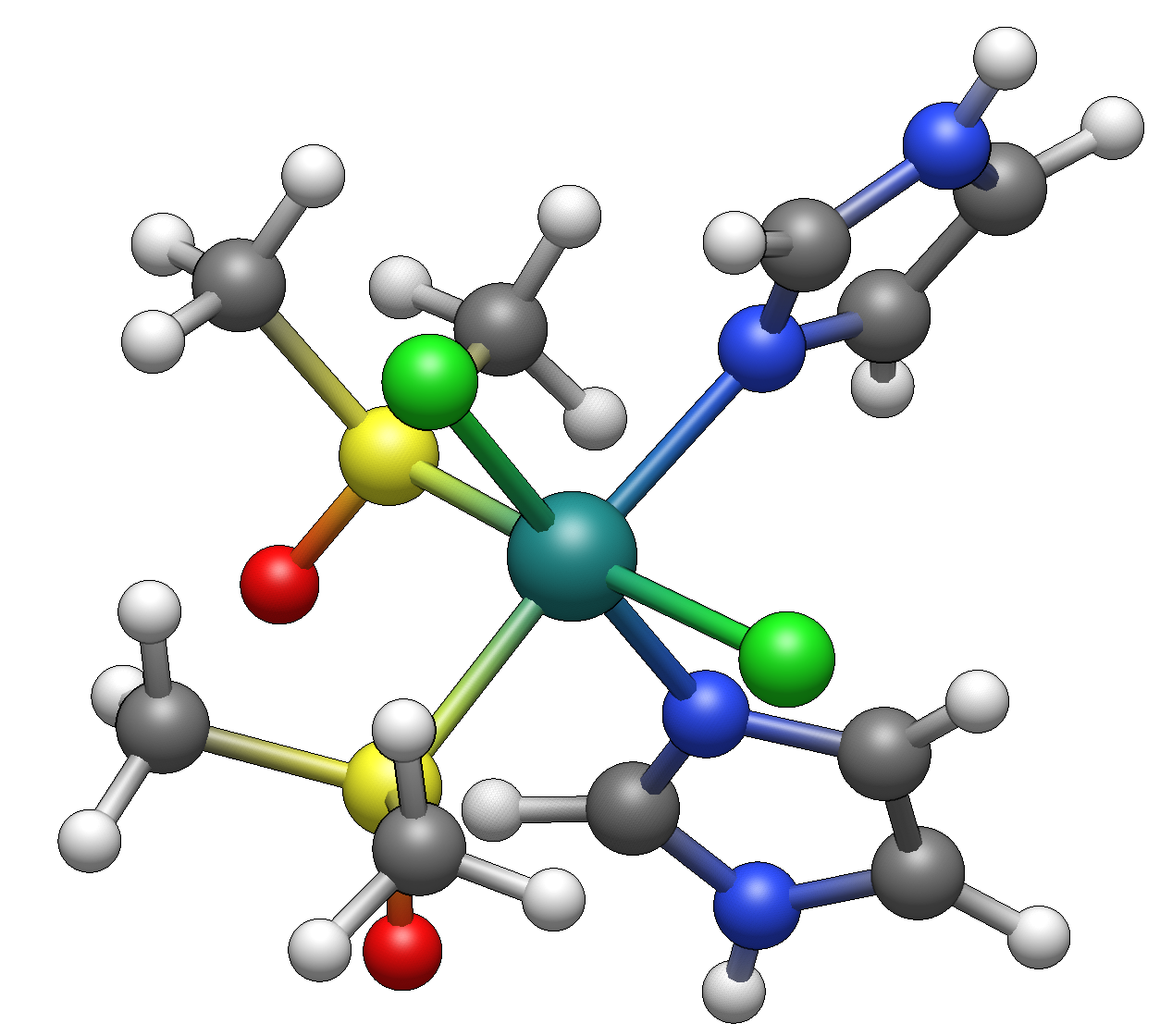}
    \caption{Molecular structure for the RuCl$_{2}$(DMSO)$_{2}$(Im)$_{2}$ Complex}
    \label{fig:my_label9}
    \end{center}
\end{figure}

 To illustrate the application of the CVS-ADC(2) implementation to medium-sized molecular systems, the method was applied to the RuCl$_{2}$(DMSO)$_{2}$(Im)$_{2}$ complex (a ruthenium-based anticancer prodrug) to compute its lowest core-excitation energy.
The optimized geometry of the complex was taken from Ref\cite{doi:10.1021/acs.inorgchem.1c02412}, and the corresponding molecular structure is shown in Figure \ref{fig:my_label9}.  
An uncontracted cc-pVDZ basis set was employed for all ligand atoms, while the dyall.v2z basis set was used for the Ru center. The resulting basis comprised 1694 spinors, of which 234 were occupied and 1460 were virtual. In the correlation treatment, 6 core electrons were kept frozen, and an SA-FNS truncation threshold of 10$^{-4}$ reduced the active space to 228 occupied and 798 virtual spinors. A Cholesky decomposition threshold of 10$^{-3}$ was used throughout, resulting in 2612 Cholesky vectors.
All calculations were carried out sequentially on a dedicated workstation equipped with dual Intel(R) Xeon(R) Gold 5315Y processors (3.20 GHz) and 2.0 TB of RAM. The computational timings were as follows: 9 h 39 min 30 s for the ground-state MP2 calculation; 1 h 28 min 18 s for ground-state density formation; 1 h 34 min 15 s for the CIS(D) calculation and state-averaged density formation in the canonical basis; and 8 h 16 min 11 s for the CVS-ADC(2) calculation. In total, the wall time required to compute the first root at the CVS-ADC(2) level was 1 day, 4 h, 18 min, and 30 s. 
The CVS-ADC(2) calculation predicts the lowest core-excitation energy at 2824.7 eV. A direct comparison with the experimental Ru L$_{3}$-edge position is not pursued here, as only a single core-excited state was computed; a meaningful comparison would require multiple states along with their associated oscillator strengths. The primary purpose of this example is therefore not to benchmark absolute excitation energies, but rather to demonstrate the computational feasibility of the present implementation for systems of this size and complexity.

\section{Conclusions}
In this work, we present an efficient implementation of the two-component relativistic CVS-ADC(2) method for core-excitation calculations, incorporating the SA-FNS and Cholesky decomposition (CD) techniques. The proposed approach significantly reduces computational cost by lowering the number of floating-point operations through SA-FNS, while CD effectively minimizes the storage requirements of two-electron integrals. As a result, the method is particularly well-suited for systems containing heavy elements. Systematic benchmarking against four-component results demonstrates the reliability and robustness of the two-component (X2CMP/X2CAMF)-based framework. Furthermore, the influence of higher-order relativistic effects on L-edge spectra has been analyzed. Convergence studies indicate that an SA-FNS truncation threshold of 10$^{-4.5}$ provides a practical and well-balanced choice for accurately describing individual excited states. Benchmark calculations of L$_{2,3}$-edge spectra for 3$d$ transition-metal compounds show that the CVS-ADC(2) method serves as a computationally efficient and reliable alternative to the non-Hermitian EOM-CC approach for reproducing experimental spectra. The practical applicability of the method is further demonstrated through a core-excitation energy calculation on a ruthenium complex, highlighting its suitability for relativistic studies of medium-sized molecular systems. Extending the present framework to higher-order variants such as CVS-ADC(2)-x is expected to further improve quantitative accuracy, and work in this direction is currently in progress.

\begin{acknowledgement}
The authors acknowledge the support from IIT Bombay, CRG, and Matrix project of DST-SERB, CSIR-India, DST-Inspire Faculty Fellowship, Prime Minister's Research Fellowship, ISRO, for financial support, IIT Bombay supercomputational facility, and C-DAC Supercomputing resources (PARAM Yuva-II, PARAM Bramha) for computational time.
\end{acknowledgement}

\begin{suppinfo}
The following file is available free of charge.
\begin{itemize}
  \item SI: The expressions for the CIS(D) 1-RDM. Comparison between the CVS-ADC(2) spectra and spectra computed using the DFT/CIS approach for the metal oxyanions and the absolute L$_{2}$-L$_{3}$ splittings and L$_{3}$/L$_{2}$ intensity ratios for both the experimental and CVS-ADC(2) spectra for metal oxyanions are provided in the SI.
\end{itemize}
\end{suppinfo}

\providecommand{\latin}[1]{#1}
\makeatletter
\providecommand{\doi}
  {\begingroup\let\do\@makeother\dospecials
  \catcode`\{=1 \catcode`\}=2 \doi@aux}
\providecommand{\doi@aux}[1]{\endgroup\texttt{#1}}
\makeatother
\providecommand*\mcitethebibliography{\thebibliography}
\csname @ifundefined\endcsname{endmcitethebibliography}  {\let\endmcitethebibliography\endthebibliography}{}

\end{document}